\documentstyle[11pt,aaspp4,amssym,flushrt]{article}

\lefthead{Cooray {\it et al.}}
\righthead{Radio Sources in Galaxy Clusters}

\begin{document}

\title{Radio Sources in Galaxy Clusters at 28.5 GHz.}

\author{Asantha R. Cooray\altaffilmark{1}, Laura Grego\altaffilmark{1,2}, William L. Holzapfel\altaffilmark{1,3}, Marshall Joy\altaffilmark{4}, John E. Carlstrom\altaffilmark{1,3}}

\altaffiltext{1}{Department of Astronomy and Astrophysics, University of Chicago, Chicago IL 60637.}
\altaffiltext{2}{Division of Mathematics, Physics, and Astronomy, California Institute of Technology, Pasadena, CA 91125.}
\altaffiltext{3}{Enrico Fermi Institute, University of Chicago, Chicago IL 60637.}
\altaffiltext{4}{Space Science Laboratory, NASA Marshall Space Flight Center, Huntsville AL 35812.}


\begin{abstract}

We present serendipitous 
observations of radio sources at 28.5 GHz (1 cm),
which resulted from 
our program to 
image thermal Sunyaev-Zeldovich (SZ) effect in 56 galaxy clusters.
In a total area of $\sim$ 0.8$^{\circ}$ sq., we find 64 radio sources
with fluxes down to $\sim$ 0.4 mJy ($> 4 \sigma$), and within 250$''$ from
the pointing centers. 
The spectral indices ($S \propto \nu^{-\alpha}$) 
of 54 sources with published low frequency
flux densities range from  
$-0.6 \lesssim \alpha \lesssim  2$ with a mean of 0.77 $\pm$ 0.06,
and a median of 0.84. 
Extending low frequency surveys of radio sources 
towards galaxy clusters CL 0016+16, Abell 665, and Abell 2218 
to 28.5 GHz, and selecting 
sources with S$_{1.4 {\rm GHz}}$ $\geq$ 7 mJy to form
an unbiased sample,
we find a mean spectral index of 0.71 $\pm$ 0.08 and a median of 0.71. 
We find 4 to 7 times more sources predicted from a 
low frequency survey in areas without galaxy clusters. This excess cannot
be accounted for by gravitational lensing of a background 
radio population by cluster potentials, indicating most of the
detected sources are associated
with galaxy clusters.
The differential source count slope, $\gamma$ $\sim$ 1.96
($dN/dS \propto S^{-\gamma}$),  
is flatter than what is expected for a nonevolving
Euclidean population ($\gamma$ = 2.5).
For the cluster Abell 2218, 
the presence of unsubtracted radio sources with S$_{28.5 {\rm GHz}}$
$\leq$ 0.5 mJy ($\sim$ 5 $\sigma$),
can only contribute to temperature 
fluctuations at a level of $\Delta T$ $\sim$ 10 to 25 $\mu$K. 
The corresponding error due to radio point source contamination
in the Hubble constant derived
through a combined analysis of 28.5 GHz 
SZ images and X-ray emission observations
ranges from 1\% to 6\%.

\end{abstract}

\keywords{galaxies: clusters: general --- radio continuum --- surveys --- techniques: interferometric}

\section{Introduction}

At present, much attention is focused on galaxy clusters 
due to the potential application
of the thermal Sunyaev-Zeldovich (SZ) effect as a cosmological tool.
Together with observations of 
X-ray emission,
a measurement of the Hubble constant can be made 
if a complete sample of galaxy clusters is used 
(see reviews by Rephaeli 1995 and Birkinshaw 1998).
Recent advances in 
interferometric tools have now allowed accurate mapping 
of the SZ decrement, producing two dimensional images which 
facilitate comparison 
of the X-ray emission and SZ effect.
The SZ effect is typically of 
arcminute scale, which is not observable with most 
interferometers designed to achieve high angular resolution. 
The exception is the Ryle Telescope which has been used successfully to image
the SZ effect at 2 cm. 
Another way to achieve the necessary beam size and sensitivity
is to use an interferometer designed for millimeter wavelengths equipped 
with low-noise centimeter-wave
 receivers. We used this approach at the Owens Valley Radio 
Observatory Millimeter Array (OVRO) and Berkeley-Illinois-Maryland Association Millimeter Array (BIMA), where we have 
now detected the SZ effect in over 20 clusters at 1 cm, with 
preliminary results given in Carlstrom {\it et al.} (1996, 1997).

The accuracy of cm-wave observations of the SZ effect can be 
limited by emission from unresolved
radio point sources towards galaxy clusters. 
The observing frequency of 28.5 GHz
was influenced four different factors: 
the large beam size required 
to be sensitive to the SZ decrement using existing interferometers,
the availability of low-noise HEMT amplifiers, atmospheric transparency,
and the expected low radio source contamination due to falling
flux density
of most radio sources with increase in frequency. 
The interferometric technique 
makes it possible to detect radio point sources with longer baselines,
which have little sensitivity to the SZ effect, 
and then remove their contribution from the short baseline data.
Though such removal will
produce point source-free SZ images, the uncertainties in removal of sources,
due to the limited signal-to-noise and imperfect coherence,
can introduce systematic 
noise. When the flux density of point sources are high, 
modeling and removal can result in systematic bias levels 
comparable to the size of the 
SZ effect. Since there are no published surveys
at 28.5 GHz,
it is not possible for us to predict accurately the number of radio sources 
expected to be present in a given cluster. In a few hours, however, 
 it is possible for us to map a cluster with sufficient
sensitivity to image
unresolved 28.5 GHz radio sources which may complicate SZ mapping. 
Clusters with no bright sources, 
are then observed for longer periods, 
$\sim$ 20 to 50 hours, to obtain adequate signal-to-noise images of the
SZ effect.

In this paper, our primary
goal is to provide information on clusters which contain radio point 
sources at 28.5 GHz.
Future 
publications will present our results on SZ detections in detail.
 Given that the cluster sample presented in this paper is incomplete 
in terms of either redshift or X-ray luminosity,  
statistical studies with this sample relating to
cluster properties should be treated with caution. 
Section 2 of this paper describes observations made with OVRO and BIMA 
arrays. The detected 1 cm radio source sample and its properties are presented
in Section 3, where we also estimate the radio source contamination
in measuring the
Hubble constant through a joint analysis of 
SZ and X-ray data by considering
galaxy cluster Abell 2218 as an example.

\section{Observations}

The observed cluster sample is presented in Table 1. The pointing centers of 
clusters were derived from existing literature  
and were checked with optical images when such images were available.
Usually, optical coordinates of the central galaxy were taken as the
pointing coordinates of a given cluster.
If there was not a clear central galaxy,
centroid coordinates from X-ray observations
were used (e.g., Ebeling {\it et al.} 1996, Ebeling {\it et al.} 1997). 
Our sample ranges in redshift
from $\sim$ 0.15 to 
0.85, with the lower limit imposed by the large angular 
scale of nearby clusters to which the interferometer
would not be sensitive,
and the 
upper limit based on the X-ray detection limit of distant clusters.  
This sample was observed at OVRO with six telescopes of the
millimeter array during summers of 1995 and 1996, 
with six telescopes of the BIMA array during summer of 1996, and 
with nine telescopes of the BIMA array during summer of 1997. 

We equipped both arrays with low-noise 
1 cm receivers, especially designed for the
detection of SZ effect. Each receiver contains a cryogenically cooled
scalar feed-horn and  HEMT
amplifier covering the frequency range 26 to 36 GHz. 
The system temperatures scaled above the atmosphere ranged from 30 to 45 K.
During the 1995 OVRO observations,
our receivers were sensitive to linear polarization. Due to 
the rotation of polarized
intensity across the sky of our calibrating sources, which
are expected to be polarized up to 10\%,
the calibration process for cluster fields observed in 1995
introduced additional uncertainties.
For long time series calibrator observations with large parallactic angle
coverage, the flux
variation can be corrected by estimating the polarization.
However, for short observations we expect an additional
5\% to 10\% uncertainty in the flux density 
of sources imaged in our
1995 cluster sample.  
We upgraded our receivers so that observations
during 1996 and 1997 detected circular polarization, which
is not subject to this effect.
For clusters that were initially observed in 1995 and were
reobserved in later years, 
we have opted to use latest data to avoid additional uncertainties.

Integration time on 
each cluster field ranged from $\sim$ 3 hours to 50 hours, 
with the short integration times on clusters where 
we happened to detect a bright radio source.
For each cluster, $\sim$ 5 minute 
observations of a secondary calibrator from the VLA 
calibrator list were interleaved
with every $\sim$ 25 minutes spent on a cluster.
Between different clusters, $\sim$ 45 to 60 minutes
were spent observing planets, with 
care taking to observe
Mars frequently since it is used as our primary flux calibrator.
The flux densities of the secondary gain and phase calibrators were 
calibrated relative to Mars. 
The brightness temperature of Mars was 
calculated
using a thermal-radiative model with an estimated uncertainty 
of 4\% (Rudy 1987). In Table 2, we present 1 cm flux densities of gain
and phase calibrators determined
through this process for the summer 1997 observations, which can
be useful for future observational programs at this wavelength. 
Some of these calibrator sources are likely to be variable at
28.5 GHz, 
but during the time scale of our 1997 observations, 2 months,
the maximum variation was found to be less than 4\%.
The uncertainties in the reported
flux densities in Table 2 are less than $\sim$ 5\%.

For the OVRO data, the MMA software package (Scoville {\it et al.} 1993) 
was used to calibrate the visibility data and then write it in UV-FITS
 format. We flagged all of the data taken when one 
antenna was shadowed by another, cluster data that was not bracketed in time
by phase calibrator data (mostly at the end or beginning of an
observation), and, rarely, data with anomalous correlations.
 We followed the same procedure for data from BIMA,
 except that MIRIAD software package (Wright \& Sault 1993) 
was used for calibration and data editing 
purposes. The image processing and CLEANing were done using  
DIFMAP (Shepherd, Pearson, \& Taylor 1994). We cleaned all fields
uniformly, based on the rms noise level.
Our automated 
mapping algorithm within DIFMAP was able to find sources with
extended structures, which when compared with low frequency 
data, such as VLA D-Array 1.4 GHz NVSS survey (Condon {\it et al.} 1996),
were confirmed for all
cases. In general, $\sim$ 2000 clean iterations 
with a low clean loop gain of 0.01 was chosen to avoid instabilities
and artifacts that can occur in fields with a large number of sources.

We looked for radio point sources in naturally weighted maps 
with visibilities
greater than 1 k$\lambda$ in BIMA data and 1.5 k$\lambda$ in OVRO data. 
Since the interferometer is less sensitive with only the long baseline data,
we obtained flux densities of detected sources in maps made with
all visibilities. Using images made with all the UV data 
also allowed us to look for
sources with extended structure.
Typical synthesized beam sizes in these images
range from 12$''$ to 30$''$.
For typical cluster and control blank fields with
no bright radio sources ($\geq$ 1 mJy), and no evident SZ decrement, 
the noise
distribution was found to be a Gaussian centered at zero.
These images did not contain any pixels with peak flux density
$\leq$ -4 $\sigma$ within 250$''$. 
The mean rms noise level for all our 
56 cluster observations is
0.24 mJy beam$^{-1}$, 
while the lowest rms noise level is 0.11 mJy beam$^{-1}$ for
BIMA observations and 0.07 mJy beam$^{-1}$ for OVRO observations.
Given the decrease in sensitivity due to the
primary beam attenuation from the image centers, 
we only report sources within 250$''$ of the pointed coordinates.
A Gaussian-noise analysis suggested that within 250$''$ from the center
in all 56 cluster fields, only 1 noise pixel is expected
at a level above 4 $\sigma$. 
Among all
56 cluster fields, there was only 
one instance where a source was 
clearly detected at a distance greater
than 250$''$ from the cluster center;
In CL 0016+16 we found a source $\sim$ 290$''$ away from the 
pointed coordinates, which is 
discussed in Carlstrom {\it et al.} (1996).

\section{Results and Discussion}

In Table 3, we report the flux densities of detected 1 cm radio sources.
When calculating these flux densities, we have corrected for the
beam response.
To determine the primary beam pattern at BIMA, the radio source 3C454.3 with
a flux density of $\sim$ 8.7 Jy at 28.5 GHz was
observed with a grid pattern of pointing offsets, and then a two dimensional
Gaussian fit was performed to the flux density values.  
A 300$''$ by 300$''$ grid with 75$''$ spacing was 
best fit by a Gaussian with a major 
axis of 386$''$, a minor axis of 380$''$ (FWHM)
and a position angle -85.31$^{\circ}$, with an uncertainty of 3$''$.  The 
rms residual from the fit was $\sim$ 0.01 Jy.  A 360$''$ by 360$''$ 
grid with 90$''$ spacing was best fit by a Gaussian with a 
major axis of 382$''$ and 
a minor axis of 379$''$, also with rms residual of 0.01 Jy.  
 Given the small 
difference between two axes and positional uncertainty of at least 
$\sim$ 5$''$ at BIMA, we have utilized a symmetrical Gaussian model with a 
380$''$ FWHM half-power point. 
At OVRO, we have made holographic measurements of 
the beam pattern and have corrected the fluxes based on the position of sources
relative to a modeled Gaussian distribution, which resulted
in a primary beam of 235$''$ (FWHM).

For our 1 cm sample, 
we searched literature for low 
frequency counterparts within 15$''$ from the 28.5 GHz radio source coordinates. 
A low frequency source was accepted
as a counterpart when the difference between our coordinates
and published coordinates was less than 
the astrometric uncertainty 
in our coordinates and the low frequency counterpart coordinates. The error
in 1 cm coordinates ranges from $\sim$ 3$''$ to 10$''$, which is equivalent
to  the
image resolution divided by the signal-to-noise with which
the source was detected.
For cluster fields with bright radio sources, 
the signal-to-noise was low due to small
integration times, producing uncertainties in position as high as $\sim$ 
10$''$.
Still, identification of such sources was easier due to their
relatively high flux densities.
For published sources, the astrometric errors ranged from sub-arcseconds, 
mostly from VLA observations, to few arcseconds. The mean
difference between our coordinates and published coordinates was
$\sim$ 6$''$. Based on Moffet \& Birkinshaw (1989), we estimated the
field density of 5 GHz radio sources towards clusters with a flux
limit of 1 mJy
is $\sim$ 25 degree$^{-2}$. Therefore, the 
probability of an unrelated radio source, with a 5 GHz flux density above
1 mJy, lying within 6$''$ is $<$ 0.5\%. 

When there is a well known
counterpart from literature coincident with the
detected 1 cm source, we have noted the commonly used name
in Table 3.
We have 
calculated the spectral index
of individual radio sources by fitting all known
flux densities,
with spectral index $\alpha$ defined as
$S \sim \nu^{-\alpha}$. In Fig.\ 1, we show a histogram of the calculated
 spectral indices of 52 sources for which we have 
found radio observations at
 other frequencies. In this plot, we have 
not included the sources 1635+6613 towards Abell 2218 and 1615-0608 towards
Abell 2163,
which are found with flux densities that peak between
1.4 GHz and 28.5 GHz (e.g., Fig.\ 2). 
These sources may indicate self-absorbed
radio cores, with spectral turnover due to free-free absorption.
Such turnovers in inverted spectra are found in Gigahertz Peaked Spectrum (GPS)
sources, though definition of GPS sources calls for peaked spectra between
0.5 and 10 GHz (De Vries, Barthel, \&  O'Dea 1997). 
The increase in turnover frequency well above
10 GHz, could be due to an increase in ambient density. 
Also, 1615-0608 towards Abell 2163 is known to be variable based on VLA 
observations by Herbig \& Birkinshaw (1994). During our observations, the flux 
density of this source did not change significantly:
we measured a flux density of 1.12 $\pm$ 0.29 mJy in 1995 (OVRO) and
0.93 $\pm$ 0.42 mJy in 1997 (BIMA) at 1 cm.
In Table 3, we report the 1995 flux density value since
tabulated VLA measurements were made closer to our 1995 observations. Abell 2163 is also known to contain one of the largest radio halo sources ever found.
We did not detect any emission from the cluster center, which is
understandable given that the halo was detected only at 1.4 GHz, with
an integrated flux density of $\sim$ 6 mJy and a steep spectral
index of $\sim$ 1.5.

In our sample we also find 3 sources with inverted spectra between
1.4 and 28.5 GHz:
0152+0102 towards Abell 267, 0952+5151 towards Zw 2701, 
and 1155+2326 towards Abell 1413.
These could 
either represent free-free emission due to starburst, or synchrotron emission
from a weak AGN, or both, with an optically thick part of a thermal
bremsstrahlung component that extends to high frequencies. Since the
inverted spectral indices are less than -2, which is the value expected for
optically thick thermal sources, it is more likely that these sources 
represent multiple non-thermal components. 
The relatively flat-spectrum ($-0.5 \lesssim
\alpha \lesssim 0.5$) sources may indicate unresolved cores and hotspots, 
and further
high resolution observations are necessary to resolve full
structure. These sources include 0152+0102 towards Abell 267 and 2201+2054 towards Abell 2409. 

In Table 3, the identification
 of a source as a central galaxy (CG) was only made when we have 
used the optical coordinates of central galaxy  
from literature as the pointing coordinates, 
and when we have detected a radio source at 1 cm within 10$''$ of the observed
coordinates. We have found 13 such
 sources, which
may well represent the radio emission associated with 
central 
cD galaxy of the cluster. 

Due to the low resolution of our observations,
most of the radio sources are unresolved, but in a few cases 
we find some evidence for extended emission. These sources include
0037+0907 and 0307+0908 towards Abell 68 (Fig.\ 3), 1716+6708 (4C +67.26) 
towards RXJ1716+6708 (Fig.\ 4), 1335+4100 (4C +41.26)
towards Abell 1763, and 1017+5934  towards Abell 959.
The nature of extended emission associated with these sources
should be further studied, and high resolution observations at several
frequencies will be helpful in this regard.  1335+4100 (4C +41.26)
towards Abell 1763 is a well studied FR II type radio source (e.g., Owen
1975).

For our sample of 52 radio sources with known flux densities at lower 
frequencies, 
a mean spectral index of 0.77 $\pm$ 0.06, and a median of 0.84 
are found. If the three sources with
inverted spectral indices are not considered, the mean and median rise to
0.85 $\pm$ 0.06 and 0.85 respectively.  
To avoid a biased estimate for the spectral index distribution,
however, we must consider counterparts at 1 cm for all
sources detected at lower frequencies. Galaxy clusters
CL 0016+16, Abell 2218 and Abell 665 have been observed
at 1.4, 4.85, 14.9 and 20.3 GHz by Moffet \& Birkinshaw (1989),
and their observations are complete to a flux density limit of 
1 mJy at 4.85 GHz.
In each of these three clusters, we selected sources in the low frequency
survey which were located within 300$''$ from the cluster center. 
We list these sources, their flux densities at 1.4 GHz,
expected flux densities at 28.5 GHz based on 1.4 and 4.85
GHz spectral index, observed flux densities at 28.5 GHz, and calculated
spectral indices between 1.4 and 28.5 GHz in Table 4.
At 28.5 GHz, we detect all sources with flux densities greater than 7 mJy
at 1.4 GHz, at a detection level greater than 
3 $\sigma$. We looked for counterparts of these sources at 28.5 GHz, 
which should form
a complete sample and not bias the spectral index distribution.
Also, given that we looked for 28.5 GHz 
counterparts only within 15$''$ of the 1.4 GHz source coordinates,
we expect all detections at a level above 3$\sigma$ to be real.  
For this sample, we find a mean spectral index of
0.71 $\pm$ 0.08, and a median of 0.71. 
The 1 cm spectral 
index distribution agrees
with that of the 6 cm mJy population with a median of 0.75 
(Donnelly {\it et al.} 1987).
However, the 1 cm distribution is steeper than the
sub-mJy and 
the $\mu$Jy populations, where medians of 0.35 
(Windhorst {\it et al.} 1993) and 
0.38 (Fomalont {\it et al.} 1991) were found at 4.85 and 8.4 GHz respectively. 
The latter sub-mJy populations have been identified with
faint blue galaxies.
Our sample could be part of the lower frequency
mJy and sub-mJy 
populations, but given the lack of detailed optical
data for most of our sources, 
we cannot exactly state the optical nature of our 28.5 GHz sample.

We compare our results with a 1.4 GHz survey by Condon, Dickey, \& Salpeter 
(1990) in areas without rich galaxy clusters, in 
order to address whether we are finding an overabundance of radio
sources at 28.5 GHz towards galaxy clusters.
They found a total of 354 radio sources, down to a flux
limit of 1.5 mJy, in a total surveyed area of about 12 square degrees. 
Seven of
these sources are thought to be associated with galaxy clusters, which 
includes
 Abell 851 (source 0942+4658 in Table 3). Ignoring this small
contamination, we calculated the expected flux densities
of the 1.4 GHz sources at 28.5 GHz based on the mean 
spectral index value of our sample.
For a spectral index of  0.71, we found 170 sources with fluxes greater than
0.4 mJy at 1 cm, 
which is the lowest 4 $\sigma$ detection limit of our observations. 
Given that the ratio of total 
area observed
by the 1.4 GHz survey and our survey is $\sim$ 15, we only expect $\sim$ 7 to
15 sources be present with flux densities greater than
0.4 mJy at 28.5 GHz, and therefore be detected in our observations. 
Given that we find 62 sources, ignoring
the inverted and unusual spectrum sources, we conclude that we are finding
at least $\sim$ 4 times more sources than usually expected. 
Given the primary beam attenuation and the nonuniformity
of flux density limit from one cluster field to another, the above ratio is
only a lower bound on the calculated ratio.
If we take these facts into account, we find that 
our sample at 28.5 GHz contains 7 times
more than one would normally 
expect based on a low frequency 
radio survey devoid of clusters. This result may have some consequences
when planning and reducing data from large field observations, such as
our planned degree-square SZ effect survey. 

The reason that we are seeing more sources might be
explained
through gravitational lensing of a background
radio population by cluster potentials. 
Our cluster sample ranges in redshift from $\sim$ 0.15 to 0.85, with a 
mean of 0.29 $\pm$ 0.02, and a median of 0.23. 
If the excess source counts are indeed an effect due to lensing,
the background population should be at a lower flux level than what
we have observed. An optimal lensing 
configuration suggests that background
sources should be at angular diameter distances twice that of the galaxy 
clusters, which are assumed to be lensing potentials.
For the range of cluster redshifts, the background source
sample should be at redshifts between $\sim$ 
0.4 and 1.4, with a mean redshift of 
$\sim$ 0.7. In terms of well known radio source samples, a 
possibility of such an unlensed population between redshifts of 0.4 and 1.4 is
the sub-mJy radio sources at 5 and 8.4 GHz (Windhorst {\it et al.} 
1993).
For simplicity, we consider a
cluster potential based on the singular isothermal sphere (SIS) model of
Schneider, Ehlers \& Falco (1992). Such a potential brightens, but dilutes the spatial distribution, 
background
sources by the magnification factor,
\begin{equation}
\mu(\theta) = \left| 1 - \frac{\theta_E}{\theta} \right|^{-1},
\end{equation}
where $\theta$ is the angle, or
distance, to radio source from cluster center,
and $\theta_E$ is the
Einstein angle, which depends on the distances to a given cluster and
background radio sources ($\theta > \theta_E$). 
The Einstein angle can be observationally determined through
optical images of clusters where background sources are lensed
into arcs and whose redshift is known. In order to estimate
reliable Einstein angles for background sources 
at redshifts around $\sim$ 0.7,
we considered two well studied clusters. In 
Abell 2218 an arc is found $\sim$ 21$''$ from the cluster center
with a measured redshift of 0.702 (Pell\'{o} {\it et al.} 1992), and
in Abell 370 an arc is found $\sim$ 36$''$, with a redshift of 0.72 (Kneib 
{\it et al.} 1994).
The 1 cm source sample ranges from $\sim$ 0 to 250$''$ in distance from
individual cluster centers, with a mean of $\sim$ 94 $\pm$ 10$''$, and
a median of 97$''$.
These values suggest a mean magnification factor of $\sim$ 1.4, suggesting
that we should expect 10 to 20 sources at 28.5 GHz towards
galaxy clusters, based on our earlier estimate of 7 to 15 sources and not
accounting for the spatial dilution due to lensing. 
It is unlikely
that lensing can account for the significant excess number of radio sources
we have detected at 28.5 GHz.
Also, VLA A-array observations at 1.4 GHz to a flux density
limit of 1 mJy
have not yet produced convincing evidence for the existence of 
gravitationally lensed radio sources, such as radio arcs,
towards galaxy clusters (e.g. Andernach {\it et al.} 1997).
An apparent detection of gravitational lensing towards clusters, based on the 
tangential orientation of radio sources, is discussed in
Bagchi \& Kapahi (1995). Recently, Smail {\it et al.} (1997) have
observed an increase in sub-mm surface flux density towards clusters Abell 
370 and CL 2244-02, which was interpreted as due to the gravitational lensing
by cluster potentials of
strongly star-forming galaxies at redshifts $\gtrsim$ 1.
Given that the number counts of our sample cannot be totally explained
as due to a lensing effect and that
we do not have enough resolution to look for alterations that might
be a result of lensing situations, we conclude that 
a large fraction of the detected 1 cm  sample 
must be associated with clusters towards which they were found.

In Fig.\ 6, we plot the source counts per solid angle
at 1 cm, which were binned in logarithmic intervals of 0.2 mJy into 
8 different bins. The solid angle for each flux bin takes into account the 
variation in sensitivity of our observations.
A large number of sources are found in the lowest bin, 
which may have a nonuniform detections
due to
the variation in noise level from one cluster field to another. 
The maximum-likelihood
fit to a power-law distribution of the observed sources,
and normalised to the 
source counts greater than 1.6 mJy, 
is $N(>S) = (59 ^{+20}_{-15}) \times (S/$mJy$)^{-0.96 \pm 0.14}$ 
in a total surveyed area of 2.5 $\times$ 10$^{-4}$ sr,
where  $N(>S)$ is an integral representation of number of
sources with flux densities greater than $S$ in mJy. 
Given
that we only looked for sources towards a 
sample of X-ray luminosity selected galaxy clusters, and that we have
not carried out 28.5 GHz observations to a uniform flux density limit in all
observed fields, the above number count-flux  relationship 
should not be treated
as true in general for all radio sources at 28.5 GHz.
However, our result may be useful when studying radio source contamination 
in planned CMB anisotropy and  SZ experiments.
The corresponding differential source count slope $\gamma$ is  $\sim$ 
1.96 ($dN/dS \propto S^{-\gamma}$).
This slope
 is similar to what is found for 6 cm mJy radio sources ($\gamma$ $\approx$ 
1.8, Donnelly {\it et al.} 1987) 
with similar flux densities as our sample, but is marginally flatter than 
sub-mJy population of radio sources ($\gamma$ $\approx$ 2.3, 
Windhorst {\it et al.} 1993). 
The flattening of the slope from the expected Euclidean value ($\gamma = 2.5$)
is likely to be due to the dependence of radio luminosity
with galaxy cluster properties, as we may be finding 
bright radio sources towards X-ray luminous clusters.

Recently, Loeb \& Refregier (1997) have suggested that
the value of the 
Hubble constant determined through a joint
analysis of SZ and X-ray data may be underestimated due to radio
point source contamination.
We address this issue based on our 28.5 GHz data and low
frequency observations towards Abell 2218, which was also
studied 
in Loeb \& Refregier (1997). 
There are 5 known sources within 300$''$ from the cluster
center (Moffet \& Birkinshaw 1989),
out of which we detect 3 (see Fig.\ 5) down to a flux density of 
$\sim$ 1 mJy.
By subtracting these three sources and using all visibilities
and a Gaussian UV taper of 0.5  
at 0.9 k$\lambda$, we find a SZ decrement 
with a signal-to-noise ratio greater than 20. The restored beam size of
this map is 110$''$ by 98$''$.
By extrapolating low frequency flux densities to 1 cm, based on the
1.4 and 4.85 GHz spectral indices, we infer an unaccounted intensity
of $\sim$ 250 Jy sr$^{-1}$. Assuming that the  
flux densities in the map at the expected location of the unsubtracted
sources are the real 28.5 GHz flux densities of the undetected
sources, we estimate an upper limit on
the unaccounted intensity of $\sim$ 590 Jy sr$^{-1}$.
The latter value is equivalent to the noise contribution in the
observed SZ decrement.
The two intensities are equivalent to $\sim$ 10 and 25 $\mu$K respectively,
which we take as the range of errors in the observed
SZ temperature decrement $\Delta T_{sz}$. 
The central temperature fluctuation
due to SZ decrement towards Abell 2218, $\Delta T_{sz}$ 
ranges from $\sim$ 0.6 to 1.1 mK, based on different $\beta$-models
to SZ morphology (see also Jones {\it et al.} 1993). The
Hubble constant, $H$, varies as $H \propto {T_{sz}}^{-2}$.
Thus, the offset in true and calculated Hubble constant, $\Delta H$, is:
\begin{equation}
\frac{\Delta H}{H} \sim \frac{2 \Delta T_{sz}}{T_{sz}}.
\end{equation}
For Abell 2218, we find that the fractional correction
to the Hubble constant from not accounting for sources 
with flux densities less than 0.5 mJy at 28.5 GHz ranges from 
$\sim$ 1\% to 6\%.
If sources with flux densities less than 0.1 mJy are not accounted for,
we estimate an upper limit on the 
offset of 2\%.
These values are in agreement with Loeb \& Refregier (1997), who suggested
that the 5 GHz
sub-mJy population (Windhorst {\it et al.} 1993)
may affect the derivation of the Hubble constant at 15 GHz by 7\% to 13\%, if
sources less than 0.1 mJy at 15 GHz not properly taking into account.
Given that the intensity of the SZ decrement 
has a spectral index of -2, and assuming a spectral
index of 0.7 for the radio source flux contribution, 
we estimate the frequency dependence  
 of the correction as $\nu^{-2.7}$. 
Thus, at 15 GHz, we also find that the Hubble constant may be
underestimated up to 13\%. The contribution from free-free emission, which 
scales as $\nu^{-0.1}$, is not expected to contribute to
underestimation of the Hubble constant
at a level more than 0.1\% at 28.5 GHz. At high frequencies ($>$ 90 GHz), 
the free-free and dust emissions, with dust scaling as
$\nu^{\beta}$, $3<\beta<4$ at 100 GHz, may become the dominant source of error.
Therefore, based on the 28.5 GHz data towards Abell 2218,
we conclude that the error in the Hubble constant through a
joint analysis of SZ data at 28.5 GHz and X-ray emission observations
 is not expected to be larger than the error introduced by the analysis
(such as $\beta$-models) 
and unknown nature of the galaxy cluster shape (oblate vs. prolate etc.), which
can amount up to 30\% (e.g. Roettiger {\it et al.} 1997).

\acknowledgments
We wish to thank the staff at OVRO and BIMA observatories for
their assistance with our observations, in particular  
J. R. Forster,
J. Lugten, S. Padin, 
R. Plambeck, S. Scott,  and D. Woody. We also thank C. Bankston and P.
Whitehouse at the MSFC for helping with the construction of the SZ receivers,
and
M. Pospieszalski for the Ka-band HEMT amplifiers. We also gratefully 
acknowledge H. Ebeling, A. Edge, H. Bohringer, S. Allen, C. Crawford,
A. Fabian, W. Voges, J. Huchra and P. Henry for providing us results from
X-ray observations of galaxy clusters prior to publication. ARC acknowledges
useful discussions with A. Fletcher on an early draft of the
paper. JEC acknowledges
support from a NSF-Young Investigator Award and the David and Lucile
Packard Foundation. Initial support to build
hardware for the SZ observations came from a NASA CDDF grant.
Radio astronomy with the OVRO millimeter array is supported
by the NSF grant AST96-13717, and astronomy with the BIMA array is
supported by the NSF grant AST96-13998.

\clearpage


\clearpage

\figcaption[fig1.ps]{Spectral-index distribution of 52 sources with 
S $>$ 0.4 mJy
at 28.5 GHz based on flux densities at lower frequencies (see Table 3). 
\label{fig1a}}

\figcaption[fig2.ps]{Radio spectrum of 1635+6619 towards Abell 2218 (source 12 of Moffet \&
Birkinshaw 1989), showing the turnover of inverted spectrum between 20.3 and 28.5 GHz.}

\figcaption[fig3.ps]{28.5 GHz map of A68 showing an extended source, 
0037+0907 north peak and 0307+0908 south peak, with
structure that corresponds with the 1.4 GHz NVSS D-Array observations of
the same cluster. The 28.5 GHz contours are at steps of 0.4 mJy (2.5 $\sigma$) 
 and shown in false color scale is the VLA D-Array NVSS 1.4 GHz map.}

\figcaption[fig4.ps]{28.5 GHz map of 1716+6711 (4C +67.26) towards RXJ1716+6708 ($z = 0.813$, Henry {\it et al.} 1997), 
where we detect the source
with a total 
flux density of 10.41 mJy and with a beam size of 24.1 by 17.8$''$ (FWHM)
at 89.8$^{\circ}$. The extended  structure of the source is clearly seen.
The 28.5 GHz contours are at 1.20 mJy (2 $\sigma$).}

\figcaption[fig5.ps]{28.5 GHz total intensity contour map of Abell 2218 which have
been plotted on to 1.4 GHz VLA D-Array NVSS observations of the same cluster.
 At 28.5 GHz, we detect 3 sources at a noise level greater than 4 $\sigma$. 
Source 1635+6619 has an unusual spectrum that peaks between 1.4 
and 28.5 GHz (Fig.\ 2). The 28.5 GHz contours are at 0.23 mJy (2 $\sigma$), 
with false color scale VLA D-Array NVSS 1.4 GHz map.}

\figcaption[fig6.ps]{The source counts at 28.5 GHz, binned in logrithmic
intervals of 0.2 mJy. These counts are fitted best with a power-law
distribution with an exponent $\sim -0.96 \pm 0.14$. The corresponding
differential source count slope $\gamma$ ($dN/dS \propto S^{-\gamma}$)
is $\sim$ 1.96.}

\figcaption[fig7.ps]{Radio sources towards Abell 520, with 28.5 GHz BIMA 1997 
total intensity contours overlaid on the 1.4 GHz VLA D-Array NVSS 
survey map. The 28.5 GHz contours are at 0.21 mJy (1 $\sigma$),
while the VLA D-Array NVSS 1.4 GHz map is represented with false color.}

\figcaption[fig8.ps]{Radio sources towards Abell 1576, with  OVRO 1996 
total intensity contours overlaid on the VLA D-Array NVSS 1.4 GHz 
map. The Double lobe source to the west is 8C 1234+634. We only detect
emission from its northern lobe, but this may be due to the rapid fall off of
the OVRO primary beam at these distances from the cluster center. The 28.5
GHz contours are at 0.22 mJy (1 $\sigma$), and the 1.4 GHz
map is shown in false color.}

\begin{figure}
\plotone{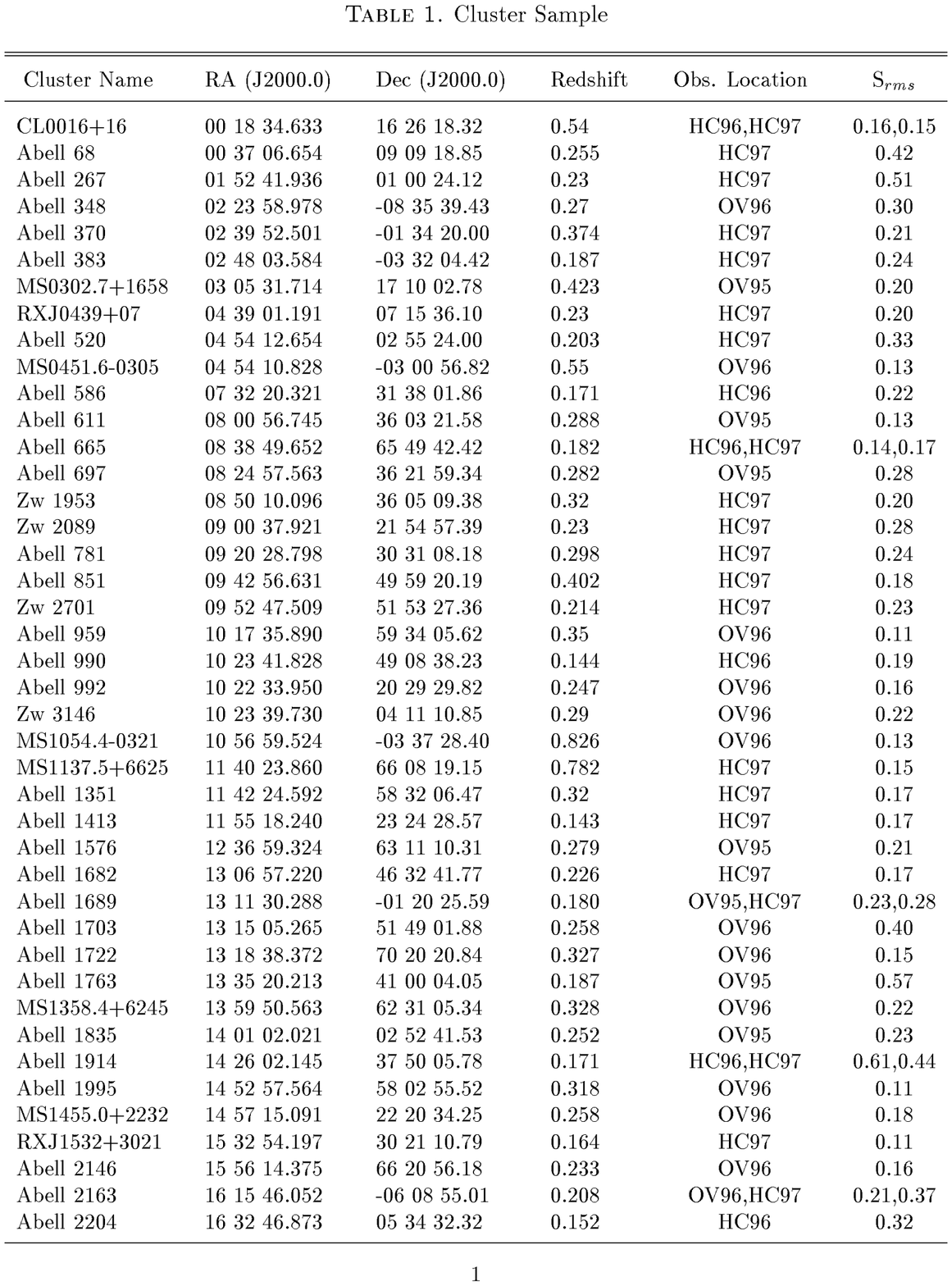}
\end{figure}

\begin{figure}
\plotone{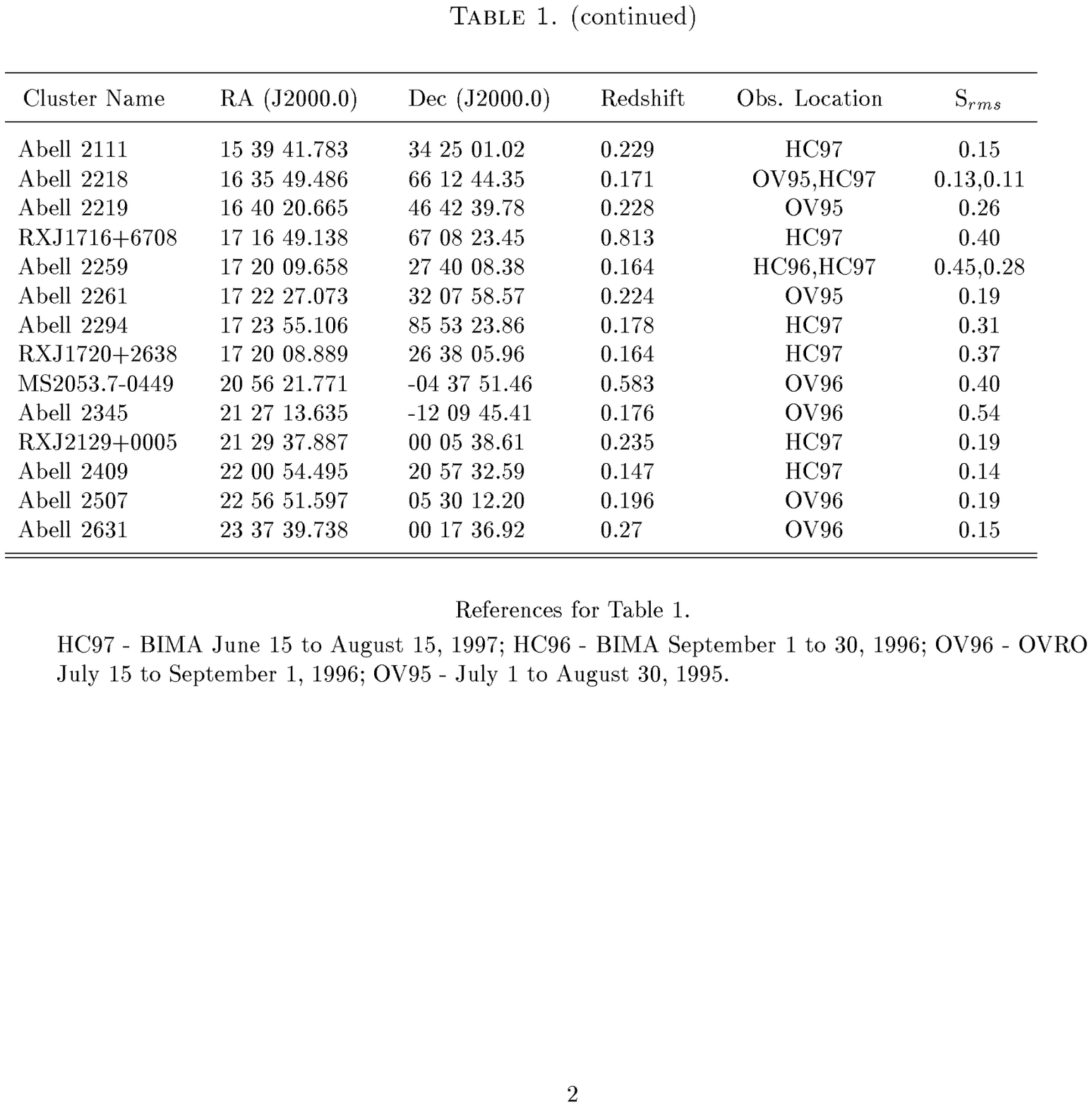}
\end{figure}

\begin{figure}
\plotone{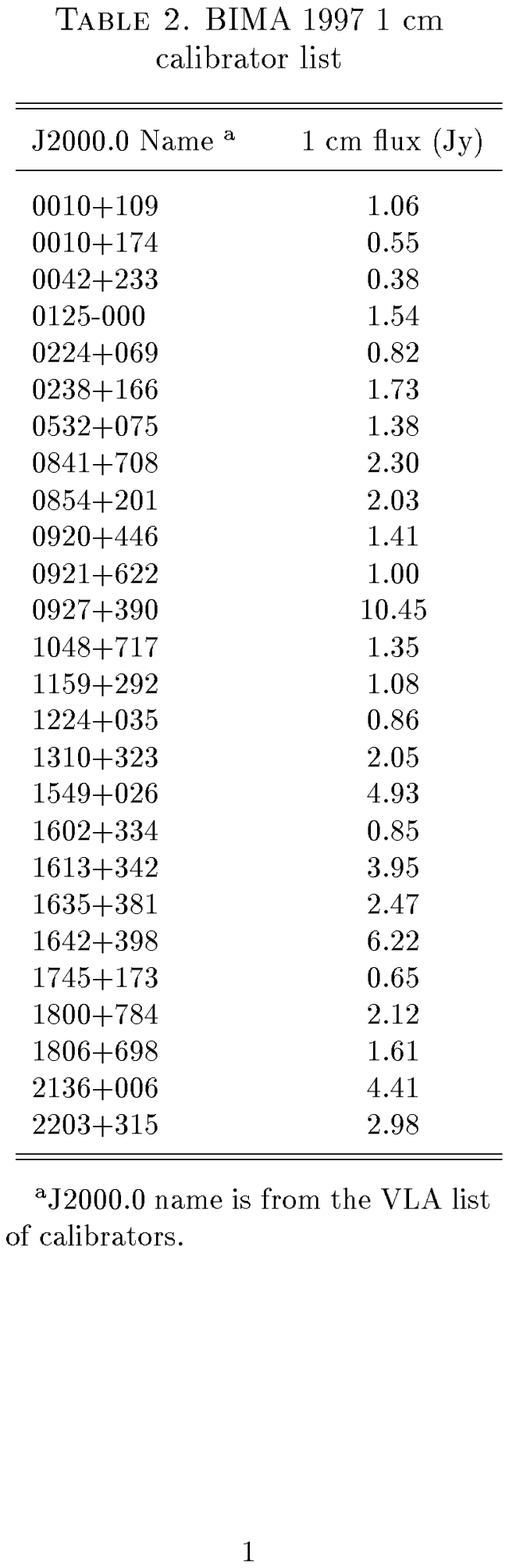}
\end{figure}
 
\begin{figure}
\plotone{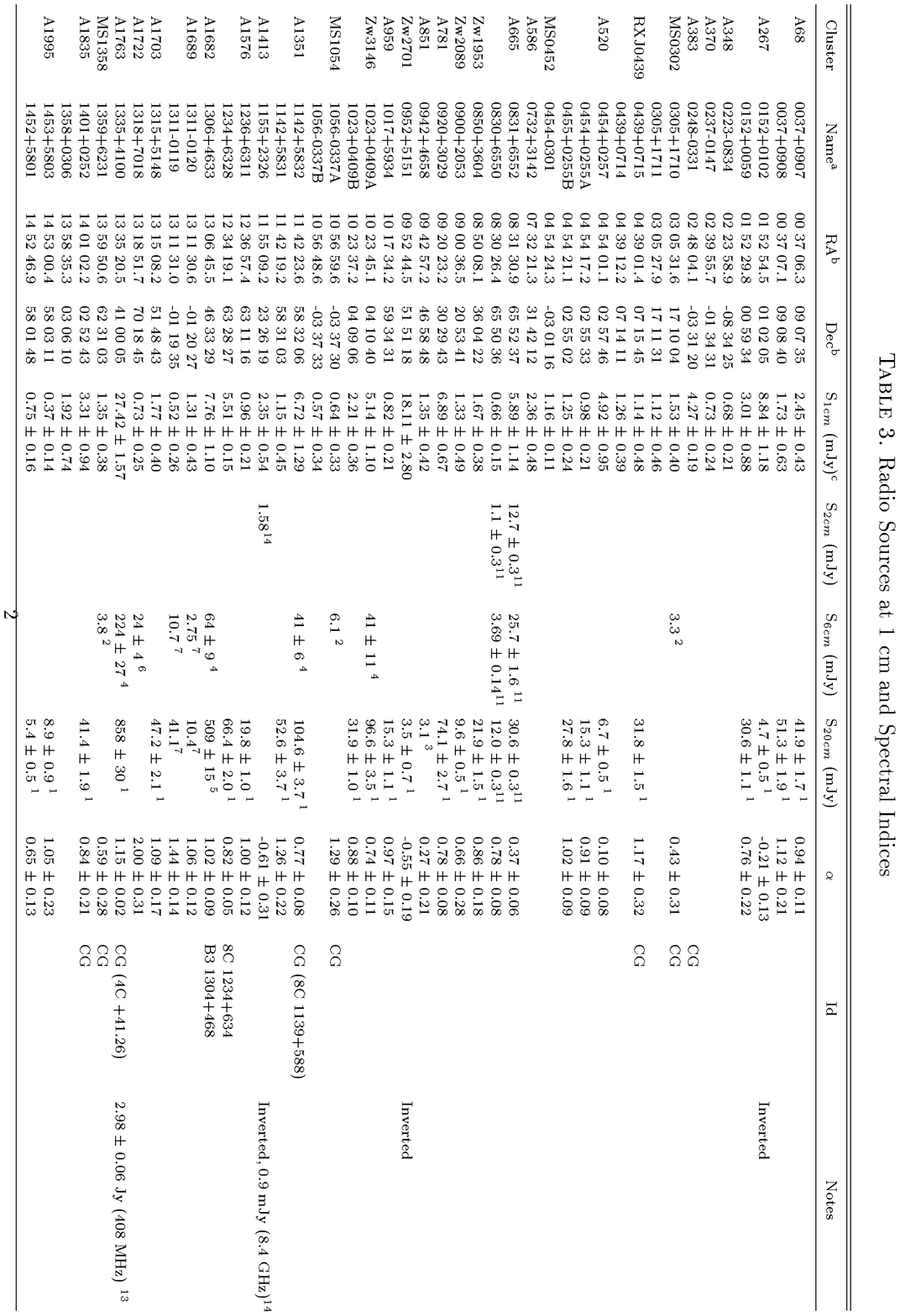}
\end{figure}
 
\begin{figure}
\plotone{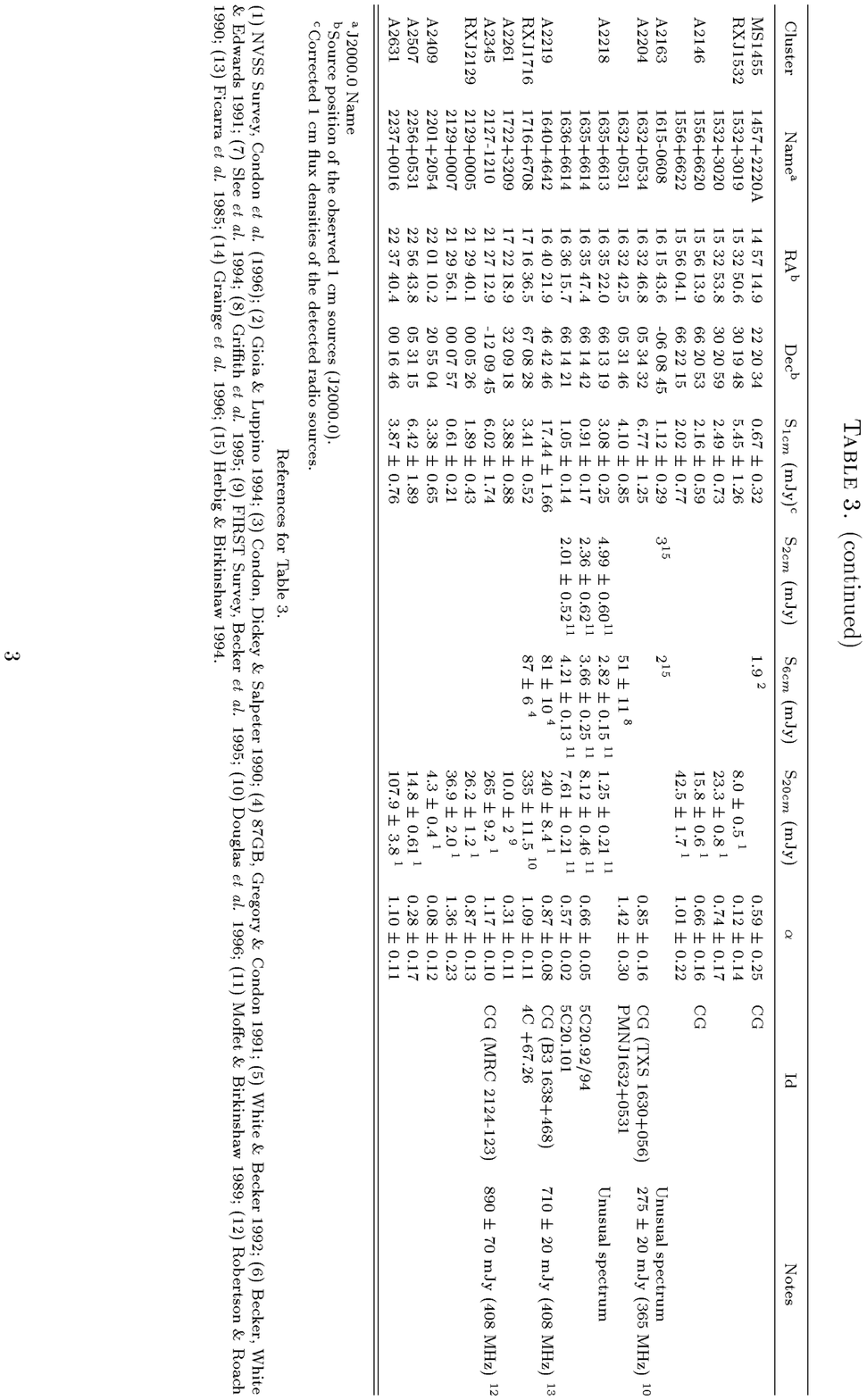}
\end{figure}
 
\begin{figure}
\plotone{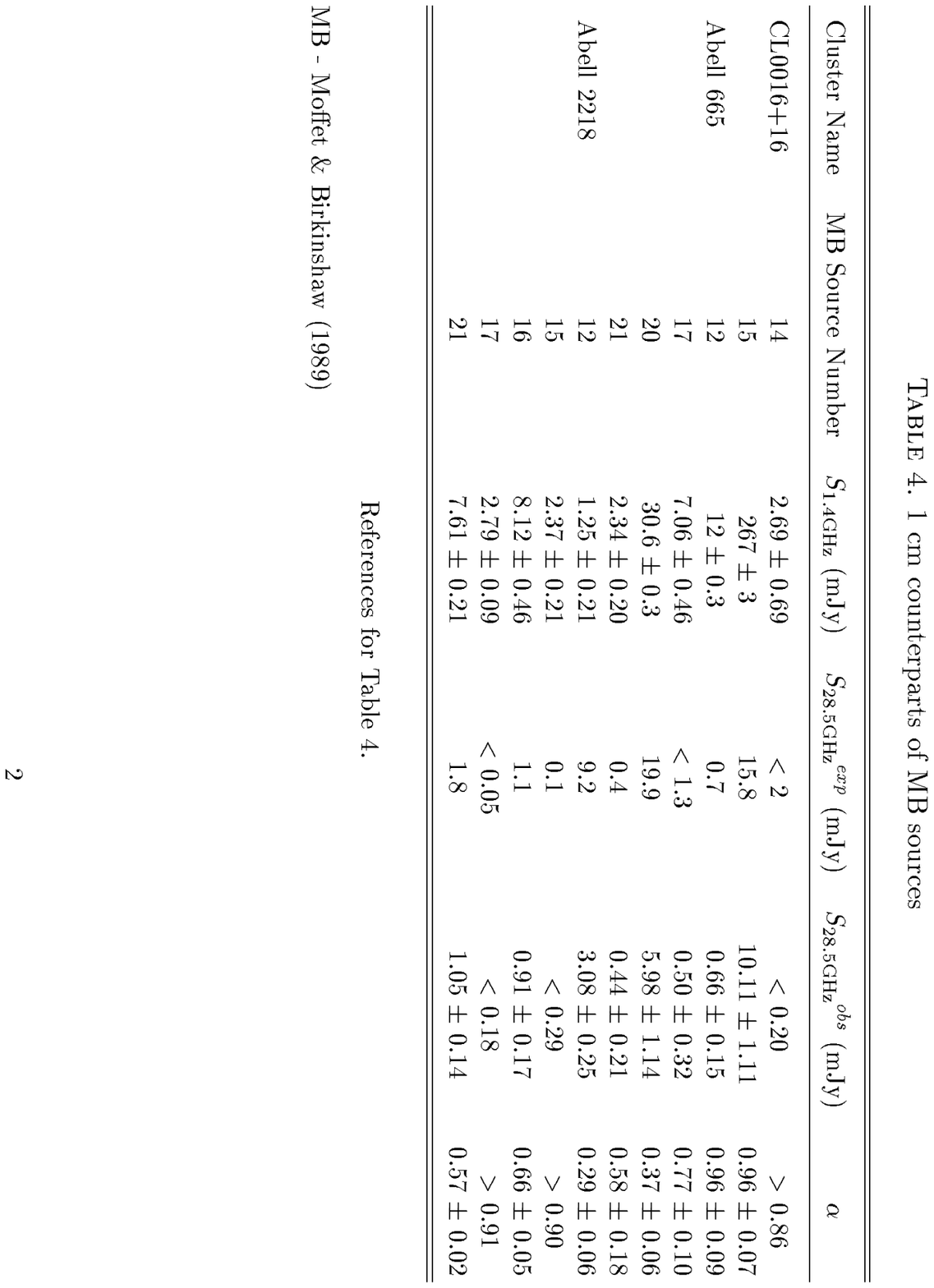}
\end{figure}


\begin{thebibliography}{}

\bibitem[Andernach et al.\ 1997]{and97} Andernach, H., Gubanov, A. G., Slee, O. B. 1997, Proc. of Observational Cosmology with the new Radio Surveys, eds.
M. Bremer, N. Jackson \& I. Perez-Fournon, Kluwer Acad. Press.

\bibitem[Bagchi et al.\ 1995]{bag95} Bagchi, J., Kapahi, V. K. 1995, JApA, 16, 131.


\bibitem[Becker et al.\ 1991]{bec91} Becker, R. H., White, R. L., Edwards, A. L. 1991, \apjs, 75, 1.

\bibitem[Becker et al.\ 1995]{bec95} Becker, R. H., White, R. L., Helfand, D. J. 1995, \apj, 450, 559.


\bibitem[Birkinshaw\ 1998]{ber98} Birkinshaw, M. 1998, submitted to Physics Reports.

\bibitem[Carlstrom et al.\ 1996]{car96} Carlstrom, J. E., Grego, L.,
 Joy, M. 1996, \apjl, 456, 75.

\bibitem[Carlstrom et al.\ 1997]{car97} Carlstrom, J. E., Grego, L.,
Holzapfel, W. L., Joy, M. 1997, Proc. of the 1996 Texas Relativistic Astrophysics Symp., ed. A. Olinto, in press.


\bibitem[Condon et al.\ 1990]{con90} Condon, J. J., Dickey, J. M., Salpeter, E. E. 1990, \aj, 99, 1071.

\bibitem[Condon et al.\ 1996]{con96} Condon, J. J., Cotton, W. D., Greisen, E. W., Yin, Q. F., Perley, P. A., Taylor, G. B., Broderick, J. J. 1996, in preparation.

\bibitem[De Vries et al. 1997]{dev97} De Vries, W. H., Barthel, P. D., O'Dea, C. P. O. 1997, A\&A, 321, 105.

\bibitem[Donnelly \ 1987]{don87} Donnelly, R. H., Partridge, R. B., Windhorst, R. A. 1987, \apj, 321,  94.

\bibitem[Douglas et al.\ 1996]{dou96} Douglas, J. N., Bas, F. N., Bozyan,
F. A., Torrence, G. W., Wolfe, C. 1996, \aj, 111, 1945.


\bibitem[Ebeling et al.\ 1996]{ebe96} Ebeling, H., Voges, W., Bohringer, H., Edge, A. C., Huchra, J. P., Briel, U. G., 1996, MNRAS, 281, 799 (see also erratum, MNRAS 283, 1103).
 
\bibitem[Ebeling et al.\ 1997]{ebe97} Ebeling, H., et al. 1997, MNRAS, submitted.


\bibitem[Ficarra et al.\ 1985]{fic85} Ficarra, A., Grueff, G., Tomassetti, G. 1985, A\&AS, 59, 255.
 
\bibitem[Fomalont\ 1991]{fom91} Fomalont, E. B., Windhorst, R. A., Kristain, J. A., Kellerman, K. I. 1991, \aj, 102, 1258.


\bibitem[Grainge\ 1996]{gra96} Grainge, K., Jones, M., Pooley, G., Saunders, R., Baker, J., Haynes, T., Edge, A. 1996, \mnras, 278, 17.

\bibitem[Gregory \& Condon \ 1991]{gre91} Gregory, P. C., Condon, J. J. 1991, \apjs, 75, 1011.


\bibitem[Griffith et al.\ 1995]{gri95} Griffith, M. R., Wright, A. E., Burke, B. F., Ekers, R. D. 1995, \apjs, 97, 347.

\bibitem[Gioia \& Luppino\ 1994]{gio94} Gioia, I. M., Luppino, G. A. 1994, \apjs, 94, 583.

\bibitem[Henry et al.\ 1997]{hen97} Henry, J. P., Gioia, I. M., Mullis, C. R., Clowe, D. I., Luppino, G. A., Boehringer, H., Briel, U. G., Voges, W., Huchra, J. P. 1997, \aj, 114, 1293.

\bibitem[Herbig and Birkinshaw\ 1994]{her94} Herbig, T., Birkinshaw, M. 1994, BAAS, 26, 1403.


\bibitem[Jones et al.\ 1993]{jon93} Jones, M., Saunders, R., Alexander, P., Birkinshaw, M., Dilon, N., Grainge, K., Hancock, S., Lasenby, A., Lefebvre, D., Pooley, G. 1993, Nature, 365, 320. 

\bibitem[Kneib\ 1994]{kne94} Kneib, J. P., Mathez, G., Fort, B., Mellier, Y., Soucail, G., Longaretti, P. Y. 1994, A\&A, 286, 701.

\bibitem[Loeb\ 1997]{loe97} Loeb, A., Refregier, A. 1997, \apjl, 476, 59.

\bibitem[Moffet \& Birkinshaw \ 1989]{mof89} Moffet, A. T., Birkinshaw, M. 1989, \aj, 
98, 1148.

\bibitem[Owen\ 1975]{owe75} Owen, F. N. 1975, \aj, 80, 263.

\bibitem[Pello\ 1992]{pel92} Pell\'{o}, R., Le Borgne, J. F., Sanahuja, B., Mathez, G., Fort, B. 1992, A\&A, 266, 6. 

\bibitem[Rephaeli \ 1995]{rep95} Rephaeli, Y. 1995, \araa, 33, 541.

\bibitem[Robertson \& Roach\ 1990]{rob90} Robertson, J. G., Roach, J. 1990,
\mnras, 247, 387.


\bibitem[Roettiger et al.\ 1997]{roe97} Roettiger, K., Stone, J. M., Mushotzky, R. F. 1997, \apj, 482, 588.

\bibitem[Rudy \ 1987]{rdu87} Rudy, D. J. 1987, Ph.D. thesis, California Institute of Technology.

\bibitem[Schnedier\ 1992]{sci92} Schneider, P., Ehlers, J., Falco, E. E. 1992,
Gravitational Lenses (New York: Springer).

\bibitem[Scoville et al.\ 1993]{sco93} Scoville, N. Z., Carlstrom, J. E., Chandler, C. J., Phillips, J. A., Scott, S. L., Tilanus, R. P. J., Wang, Z. 1993, \pasp, 105, 1482.

\bibitem[Shepard et al.\ 1994]{she94} Shepherd, M. C., Pearson, T. J., Taylor, G. B. 1994, BAAS, 26, 987.

\bibitem[Slee et al.\ 1994]{sle94} Slee, O. B., Roy, A. L., Savage. A. 1994,
Australian Journal of Physics, 47, 145.

\bibitem[Smail et al.\ 1997]{sma97} Smail, I., Ivison, R. J., Blain, A. W. 1997, \apjl, in press (also astro-ph/9708135).



\bibitem[White \& Becker \ 1992]{whi92} White, R. L., Becker, R. H. 1992, \apjs, 79, 331.

\bibitem[Windhorst\ 1993]{win93} Windhorst, R. A., Fomalont, E. B., Partridge, R. B., Lowenthal, J. D. 1993, \apj, 405, 498.

\bibitem[Wright \& Sault\ 1993]{wri93} Wright, M. C. H., Sault, R. J. 1993, \apj, 402, 546.

\end{thebibliography}
\end{document}